\documentclass[aps,prl,reprint]{revtex4-1}
\usepackage{color}
\definecolor{darkblue}{RGB}{0,0,128}
\usepackage{graphicx}
\usepackage{dcolumn}
\usepackage{bm}
\usepackage[breaklinks, colorlinks=true, citecolor=darkblue, urlcolor=darkblue, linkcolor=black]{hyperref}

\begin{document}


\title{Accurate Measurement of the Gap of Graphene/hBN Moir\'e Superlattice through Photocurrent Spectroscopy}

\author{Tianyi Han$^1$, Jixiang Yang$^1$, Qihang Zhang$^1$, Lei Wang$^{2,3,4}$,\\Kenji Watanabe$^5$, Takashi Taniguchi$^6$, Paul L. McEuen$^{2,3}$, Long Ju$^1$}
\email{longju@mit.edu}
\affiliation{$^1$Department of Physics, Massachusetts Institute of Technology, Cambridge, MA 02139, USA\\
$^2$Kavli Institute at Cornell for Nanoscale Science, Ithaca, NY 14853, USA\\
$^3$Laboratory of Atomic and Solid State Physics, Cornell University, Ithaca, NY 14853, USA\\
$^4$National Laboratory of Solid-State Microstructures, School of Physics and Collaborative Innovation Center of Advanced Microstructures, Nanjing University, Nanjing 210093, China\\
$^5$Research Center for Functional Materials, National Institute for Materials Science, 1-1 Namiki, Tsukuba 305-0044, Japan\\
$^6$International Center for Materials Nanoarchitectonics, National Institute for Materials Science, 1-1 Namiki, Tsukuba 305-0044, Japan}






\date{\today}

\begin{abstract}
Monolayer graphene aligned with hexagonal boron nitride (hBN) develops a gap at the charge neutrality point (CNP). This gap has previously been extensively studied by electrical transport through thermal activation measurements. Here, we report the determination of the gap size at CNP of graphene/hBN superlattice through photocurrent spectroscopy study. We demonstrate two distinct measurement approaches to extract the gap size. A maximum of $\sim$ 14 meV gap is observed for devices with a twist angle of less than 1 degree. This value is significantly smaller than that obtained from thermal activation measurements, yet larger than the theoretically predicted single-particle gap. Our results suggest that lattice relaxation and moderate electron-electron interaction effects may enhance the CNP gap in hBN/graphene superlattice.
\end{abstract}

\maketitle


Moir\'e superlattice composed of two-dimensional (2D) materials can be generated by either a twist between adjacent layers or a lattice mismatch between different materials. The resulting moir\'e potential could dramatically change the electronic and optical properties of 2D materials\cite{hunt2013massive,ponomarenko2013cloning,woods2014commensurate,ribeiro2018twistable,yankowitz2018dynamic,finney2019tunable,wang2015evidence,shi2014gate,yankowitz2012emergence,jin2019observation,jin2019identification,eckmann2013raman}. More recently, 2D moir\'e superlattices have emerged as a new platform to engineer topology and electron correlations. The periodicity of a moir\'e superlattice determines the Coulomb repulsion energy, while the moir\'e potential quenches the kinetic energy (with the help from external electric and magnetic fields)--bringing the system to a correlated regime.  Experiments have demonstrated correlated insulator\cite{cao2018correlated,chen2019evidence,liu2020tunable,cao2020tunable,tang2020simulation,regan2020mott,wang2020correlated}, superconductivity\cite{cao2018unconventional,yankowitz2019tuning,lu2019superconductors,chen2019signatures} and ferromagnetism\cite{sharpe2019emergent,serlin2020intrinsic,chen2020tunable} in a variety of 2D moir\'e superlattices. The accurate quantification of the moir\'e potential is crucial to better understand the underlying physics of these emerging phenomena, and to utilize moir\'e superlattices to design and realize more exotic electronic ground states. For example, people found that when twisted bilayer graphene is further aligned with hBN, ferromagnetism and quantized anomalous Hall effect emerge\cite{sharpe2019emergent,serlin2020intrinsic}, where the estimation of the moir\'e potential at twisted bilayer graphene/hBN interface is the key to account for the $C_2$ symmetry breaking of graphene. However, spectroscopy studies of these systems have so far remained limited.

Among various 2D moir\'e superlattices, monolayer graphene/hBN superlattice is the first system that has been extensively studied both theoretically and experimentally\cite{yankowitz2012emergence,hunt2013massive,ponomarenko2013cloning,woods2014commensurate,eckmann2013raman,giovannetti2007substrate,sachs2011adhesion,kindermann2012zero,wallbank2013generic,song2013electron,shi2014gate,jung2014ab,moon2014electronic,jung2015origin,jung2017moire,ribeiro2018twistable,yankowitz2018dynamic,finney2019tunable}. The boron and nitrogen atoms break the sublattice symmetry in graphene, which opens local band gaps at the charge neutrality point (CNP) of graphene\cite{giovannetti2007substrate}. However, it is calculated that such a local gap is nearly averaged out by different stacking order regions in a global moir\'e superlattice\cite{sachs2011adhesion}. On the other hand, the moir\'e potential breaks the inversion symmetry, resulting in a global spectral gap opening at the CNP\cite{kindermann2012zero,moon2014electronic,jung2017moire}. The periodicity of the moir\'e superlattice can modulate the gap size, where the maximum value of the moir\'e period is around 14 nm for zero-degree twist between the graphene and hBN lattice\cite{ribeiro2018twistable}. Electron transport measurements of this gap size via thermal activation yield a maximum value in the range of 30--40~meV\cite{hunt2013massive,woods2014commensurate,wang2015evidence,ribeiro2018twistable,finney2019tunable}, which is 4--5 times bigger than the prediction from the single-particle model($\sim$7~meV with the consideration of lattice relaxation\cite{jung2015origin}). The deviation is attributed to possible enhancement by electron-electron interactions\cite{hunt2013massive,song2013electron,bokdam2014band}. Compared with various electron transport measurements, spectroscopic study of this gap remains scarce. It was claimed that a 38~meV gap exists at the CNP by using infrared transmission spectroscopy measurement\cite{chen2014observation}. However, their method of extrapolation from inter Landau level transitions at high magnetic field is indirect. Together with the neglected variation of Fermi velocity at different energy, such a measurement suffers from many experimental uncertainties. Recently, a planer tunneling spectroscopy experiment observed a gap size of around 16~meV\cite{kim2018accurate}, significantly smaller than that in all other experiments.

Here we report the systematic study of the CNP gap in graphene/hBN superlattice by photocurrent spectroscopy. We demonstrate two ways to determine the gap size, one is based on the onset energy of interband optical transition at zero magnetic field, while the other one is based on the splitting of Landau level (LL) transition peaks at finite magnetic field. Both experimental approaches show consistent results, where the maximum gap size is around 14 meV. This value is consistent with the tunneling spectroscopy experiment, and substantially smaller than most transport results.

Fig. 1(a) shows an optical image of the device (device C). We fabricated high quality monolayer graphene Corbino devices encapsulated between two flakes of hBN (except for device A, which is a van der Pauw structure device. See Supplemental Material\cite{supplemental} and Ref. \cite{wang2013one} for device fabrication details).The graphene layer forms a small twist angle $\theta$ with either the top or bottom hBN flake. The layered stack is placed on either a graphite local bottom gate or a SiO$_2$/Si global bottom gate. Fig. 1(c) shows the Landau fan diagram of our Corbino device C. Pronounced LLs develop at magnetic field less than 0.5~T, demonstrating the device is of high quality. Satellite conductance dips arising from the moir\'e gap are observed at V$_g\sim\pm$6~V, which develop into separate sets of LLs. Based on the LL degeneracy and the positions of satellite conductance dips, we can extract a moir\'e wavelength of $\sim$14nm and a twist angle of $\sim$0.1$^{\circ}$ for this device. We use the same method to extract the moir\'e wavelength and twist angle in other devices. See details in Supplemental Material\cite{supplemental}.

\begin{figure}
\includegraphics[width=1\columnwidth]{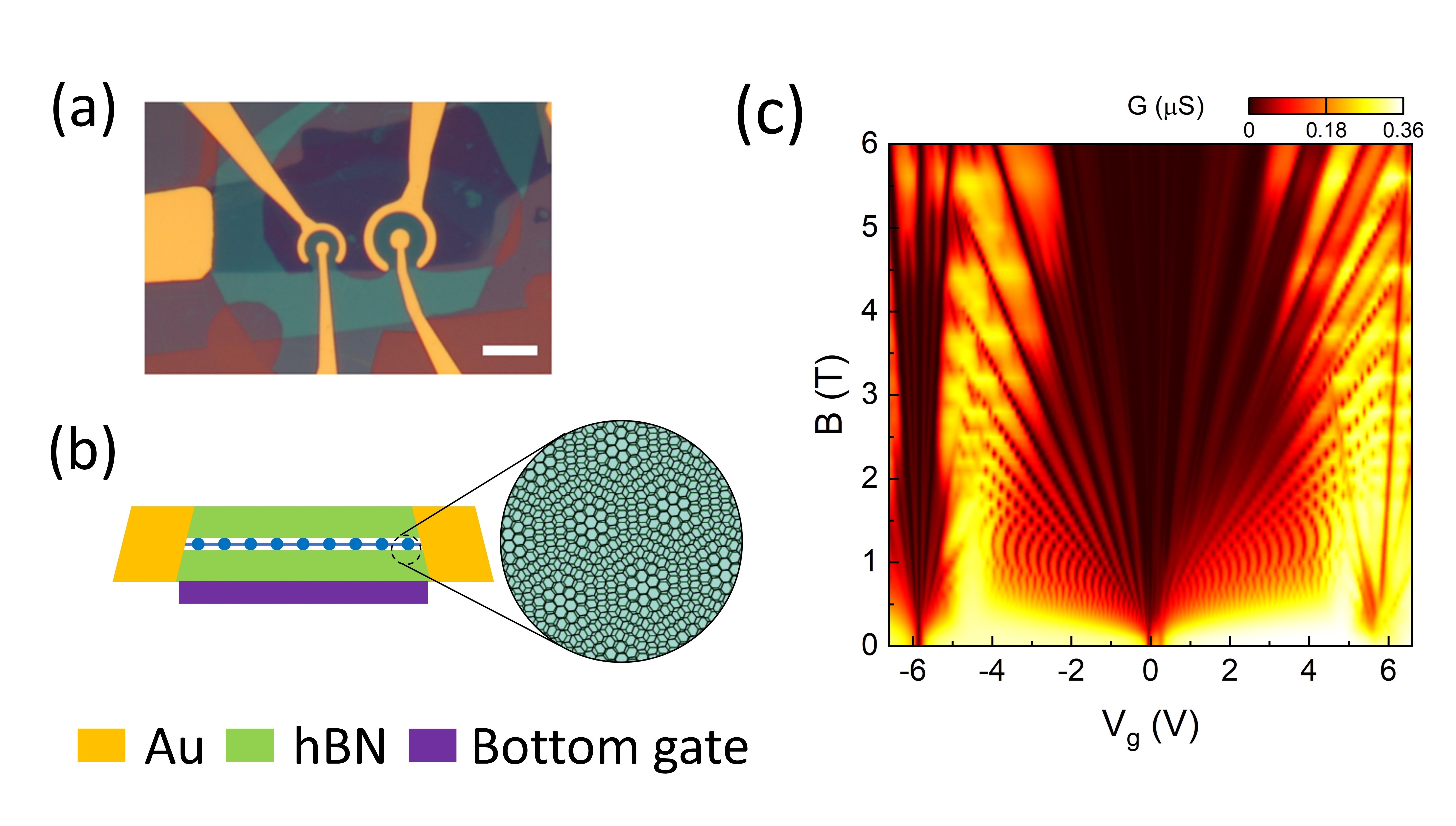}
\caption{\label{fig1}Device structure and transport characterization. (a) Optical image of a Corbino geometry device. Scale bar: $10\mu m$. (b) Schematic of the device structure: monolayer graphene sandwiched by hBN, with either a SiO$_2$/Si or graphite bottom gate. The moir\'e superlattice exists between monolayer graphene and either top or bottom hBN flake. (c) Characteristic Landau fan diagram of the monolayer graphene Corbino device C with moir\'e potential. Satellite conductance dips emerge at around V$_g$=$\pm$6~V.}
\end{figure}

We then use Fourier transform infrared (FTIR) photocurrent spectroscopy\cite{ju2017tunable,ju2020unconventional} to measure the optical absorption spectrum (see measurement details in Supplemental Material\cite{supplemental}). In the quantum Hall regime, the edge current in devices would contribute to a large background signal, which would obscure our signal from inter-LL optical transitions. Thus, we fabricate the graphene/hBN devices in a Corbino geometry as shown in Fig. 1(a) to eliminate the edge channel, and only collect photocurrent signal from the bulk. We mainly show the photocurrent data of three high quality devices in the main text, i.e. device A, B and C. The data of other devices are included in the Supplemental Material\cite{supplemental}. All data was collected at a base temperature of 2 K.

\onecolumngrid

\begin{figure}[!hbp]
\includegraphics[width=0.97\columnwidth]{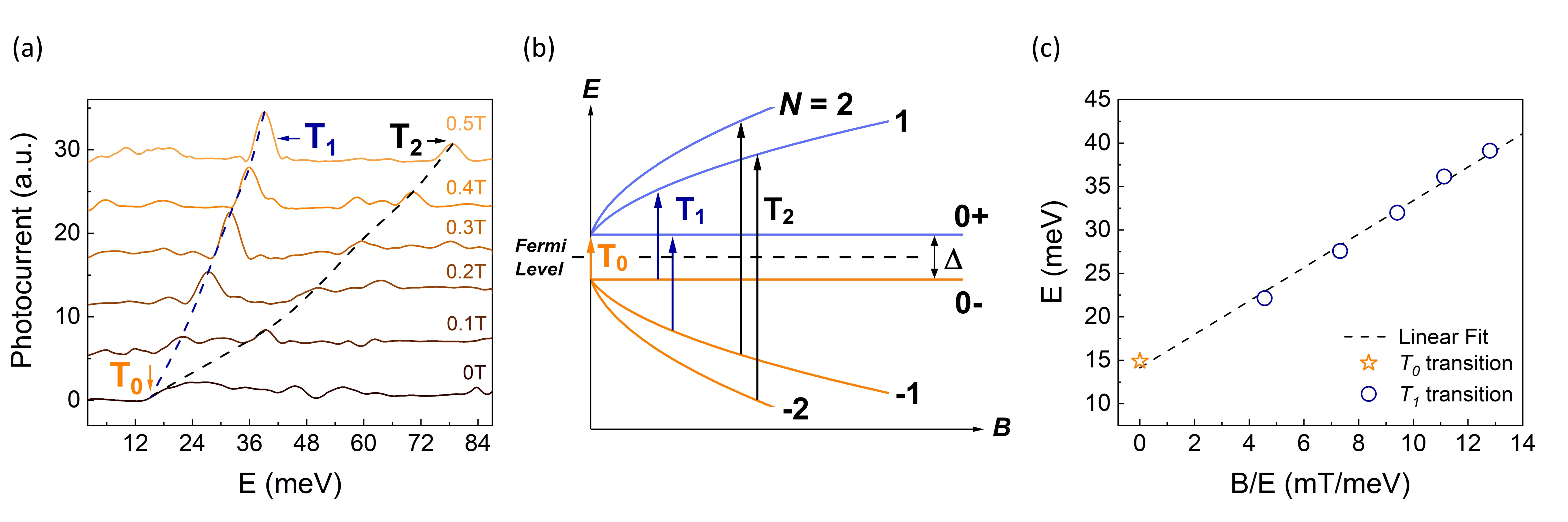}
\caption{\label{fig2} Photocurrent spectra with the Fermi level located at the CNP. (a) Photocurrent spectra at magnetic field B = 0-0.5 T. At B = 0 T, the optical transition has a threshold energy of $T_0$, which corresponds to the bandgap $\Delta$. When B increases, LLs are formed and inter-LL transitions labelled as $T_1$ and $T_2$ emerge from the continuous spectrum at 0 T. Spectra are shifted vertically for clarity. Dashed curves are guide lines of the $T_1$ and $T_2$ transitions. (b) Schematic illustrating the inter-LL transition $T_1$ and $T_2$ that are allowed by the optical selection rule $\Delta |N|=\pm 1$. At the low magnetic field limit, they both converge to $T_0$ transition energy $\Delta$. (c) Extraction of $\Delta$ from the fitting of $T_0$ (14.9 meV marked by star symbol in orange) and inter-LL transition $T_1$ by using $E=2e\hbar{v_F}^2B/E+\Delta$.}
\end{figure}
\twocolumngrid

We demonstrate two approaches of determining the gap size at the CNP in graphene/hBN superlattice through the photocurrent spectroscopy. The first one starts with photocurrent spectroscopy at zero magnetic field. By gating we placed the Fermi level at the CNP. With a gap opened at the CNP, interband transitions are forbidden at low energy while they start to appear at a threshold energy. This threshold energy provides a first measurement of the band gap size which can be clearly resolved in Fig. 2(a). The device used here has a moir\'e wavelength of 10.4 nm and a corresponding twist angle of 1.0$^{\circ}$ (device A). Due to thermal and disorder broadening, the expected step function of optical absorption is smeared. So, we used the mid-point of the rising slope in spectrum to determine $T_0$= 14.9 meV. This gap value is 2--3 times smaller than the previous reported value in transport measurement\cite{hunt2013massive,woods2014commensurate,wang2015evidence,ribeiro2018twistable,finney2019tunable}. We further verify our results by applying a small out-of-plane B field to the device while fixing the Fermi level at the CNP. LLs emerge at a magnetic field as low as 0.1 T, which is signified by the sharp inter-LL optical transitions $T_1$ and $T_2$ as traced by dashed curves in Fig. 2(a). We illustrate the allowed inter-LL optical transitions in the Fig. 2(b). The selection rule for inter-LL transitions is $\Delta |N|=\pm 1$, where $N$ is the LL index. When the Fermi level is in the CNP gap, allowed LL transitions are labelled in the diagram as $T_1$, $T_2$, etc. In this picture, the energy of the $T_1$ transition should be described by a simple model $E=\sqrt{2e\hbar{v_F}^2B+(\Delta/2)^2}+\Delta/2$. Through a simple transformation we obtain: $E=2e\hbar{v_F}^2B/E+\Delta$. We tested this model by summarizing experimentally extracted $T_0$ and $T_1$ as shown in Fig. 2(c). All the data can be well fitted with a gap value $\Delta$=14.0$\pm$0.6 meV and Fermi velocity $v_F\sim$1.21($\pm$0.02)$\times 10^6$ m/s. We also fit the $T_0$ and $T_2$ with the formula of the $T_2$ transition $E_{T_2}=\sqrt{2e\hbar{v_F}^2B+(\Delta/2)^2}+\sqrt{2\times (2e\hbar{v_F}^2B)+(\Delta/2)^2}$, and extract a gap value $\Delta$=15.3$\pm$1.2 meV and a Fermi velocity $v_F\sim$1.24($\pm$0.01)$\times 10^6$ m/s, which is on the same order of magnitude with the results from the $T_1$ transition. Details of the fitting of $T_2$ transition are included in the Supplemental Material\cite{supplemental}. As we have a high-quality device and a large signal-to-noise ratio of our FTIR photocurrent spectrum, we can directly resolve the optical absorption information at B $<$ 0.5 T and measure the CNP gap size.

\begin{figure}
\includegraphics[width=1\columnwidth]{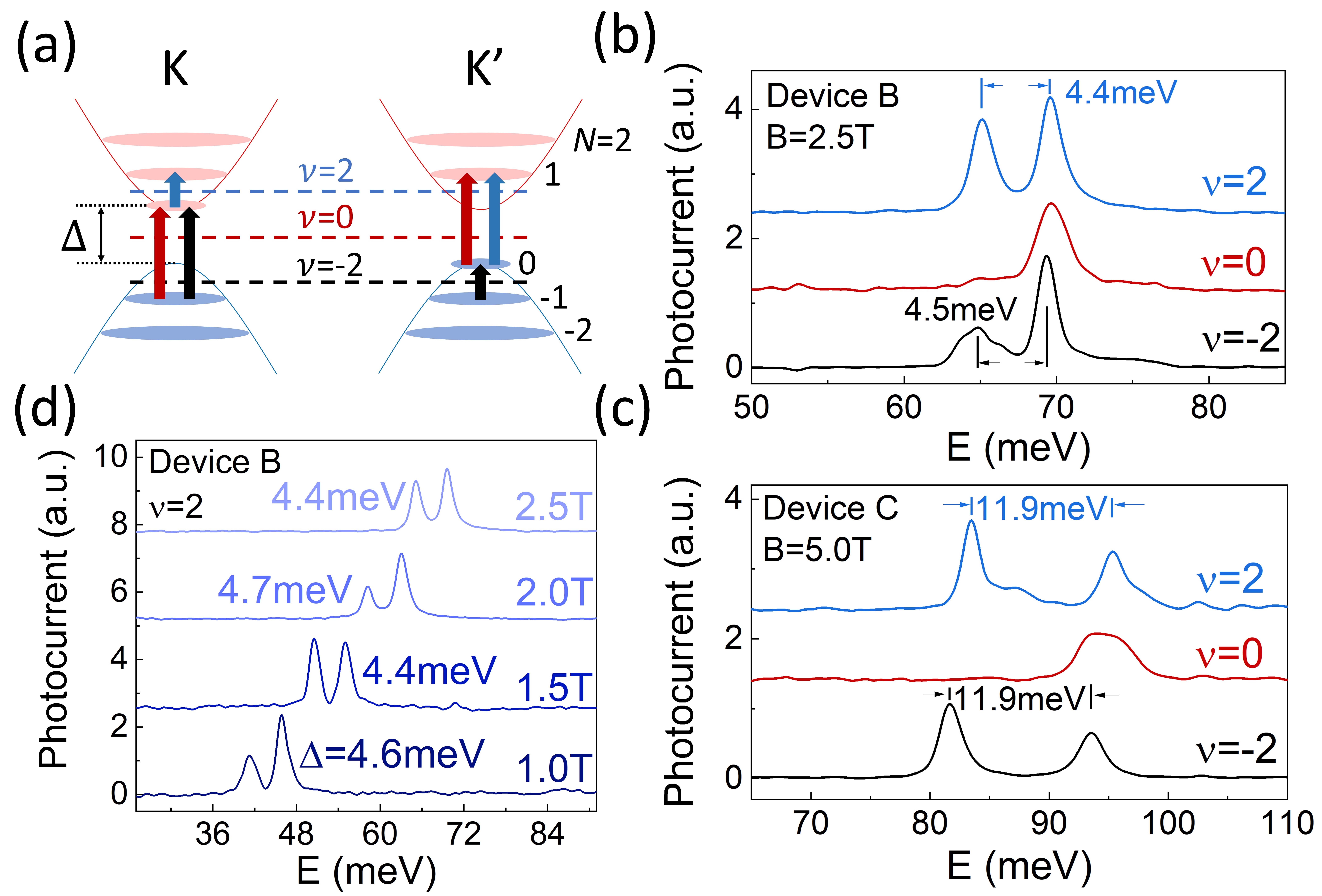}
\caption{\label{fig3}Extraction of $\Delta$ from inter-LL transitions at different filling factors. (a) Schematic showing allowed inter-LL transitions in the K and K$^{\prime}$ valleys of graphene when the Fermi level is adjusted to filling factor $\nu$ = -2 (black dashed line and arrows), $\nu$ = 0 (red) and $\nu$ = 2 (blue). The zeroth LL is located at the bottom of the conduction band (top of the valence band) in the K (K$^{\prime}$) valley --lifting the valley degeneracy of inter-LL transitions. (b-c) Photocurrent spectra of device B at 2.5 T and device C at 5.0 T. In both cases, the two allowed inter-LL transitions remain degenerate at $\nu$ = 0, resulting in a single absorption peak. At both $\nu$ = 2 $\&$ -2, the two allowed transitions are split by the gap energy $\Delta$. Transitions at $\nu$ = 2, 0, -2 correspond to blue, red and black arrows in (a) respectively. (d) Photocurrent spectra at $\nu$ = 2 for device B as a function of magnetic field from 1.0 T to 2.5 T. The splitting between two inter-LL transitions in all spectra are close to 4.5 meV as labelled.}
\end{figure}

We adopted a second approach to extract the gap size at the CNP through inter-LL transitions at finite magnetic fields. The idea is illustrated in the schematic in Fig. 3(a). The zeroth LL of the massive Dirac fermion is located at the bottom of conduction band in the K valley and the top of valence band at the K$^{\prime}$ valley. This valley polarization does not result in significant peak splitting in optical absorption when the Fermi level is at the CNP and the filling factor at $\nu$ = 0, since the inter-LL transition (-1$\rightarrow$ 0, K) and (0$\rightarrow$ 1, K$^{\prime}$) as indicated by red arrows are mostly degenerate. However, when the Fermi level is placed between LL $N$ = 1 and 0 in the K valley or $N$ = -1 and 0 in the K$^{\prime}$ valley, Pauli blocking forbids one transition that was allowed at $\nu$ = 0 and enables another inter-LL transition. This change in optical transitions is illustrated by blue (black) arrows when the filling factor equals +2 (-2) respectively in Fig. 3(a). Therefore, the optical absorption spectrum features two transition peaks that are separate in energy by $\Delta$ -- allowing us to determine the CNP gap independently from the first experimental approach. We utilized the bottom gate to tune the Fermi level and obtained typical photocurrent spectra for different filling factors as shown in Fig. 3(b) $\&$ (c) for device B and C. At $\nu$ = 0, the photocurrent spectrum features a single peak in the range of $T_1$ transitions. The relatively wide peak width in Fig. 3(c) indicates a small splitting due to electron-hole asymmetry of the band structure, which does not affect our conclusion. At both $\nu$ = +2 and -2, the spectrum features two distinct peaks that are widely separated by an energy much bigger than the peak width. This result agrees with our expectation based on Fig. 3(a), and allowed us to extract a CNP gap size of 4.5 meV for device B and 11.9 meV for device C. We further explore the peak splitting as a function of magnetic field for device B as shown in Fig. 3(d). The filling factor at all B fields are fixed at $\nu$ = +2. The splitting remains at $\sim$4.5 meV with small variations. This consistent value in a wide range of magnetic field proves the validity of our model and experimental methodology.

In Fig. 4 we summarize the gap sizes measured in all devices through the photocurrent spectroscopy with the two different approaches. For devices with a small twist angle, we are able to extract the moir\'e superlattice period (as listed in the x axis in blue shaded area in Fig. 4) and the twist angle. We also measured devices with bigger twist angles than 1$^{\circ}$, but due to the limited charge density that can be achieved by gating, we do not know exactly what their twist angles are. The band gap sizes extracted from these devices are listed in pink shaded area and are all smaller than the ones with twist angles less than 1$^{\circ}$. This trend is phenomenologically consistent with the single-particle picture. We note that the single data point at $\sim$ 1$^{\circ}$ with a relatively larger error bar (denoted by the orange circle in Fig. 4) is extracted from the onset energy of interband transition and the fitting, which may be less accurate than other data points in Fig. 4 that are directly extracted from the energy splitting of the LL transition spectra.

\begin{figure}
\includegraphics[width=1\columnwidth]{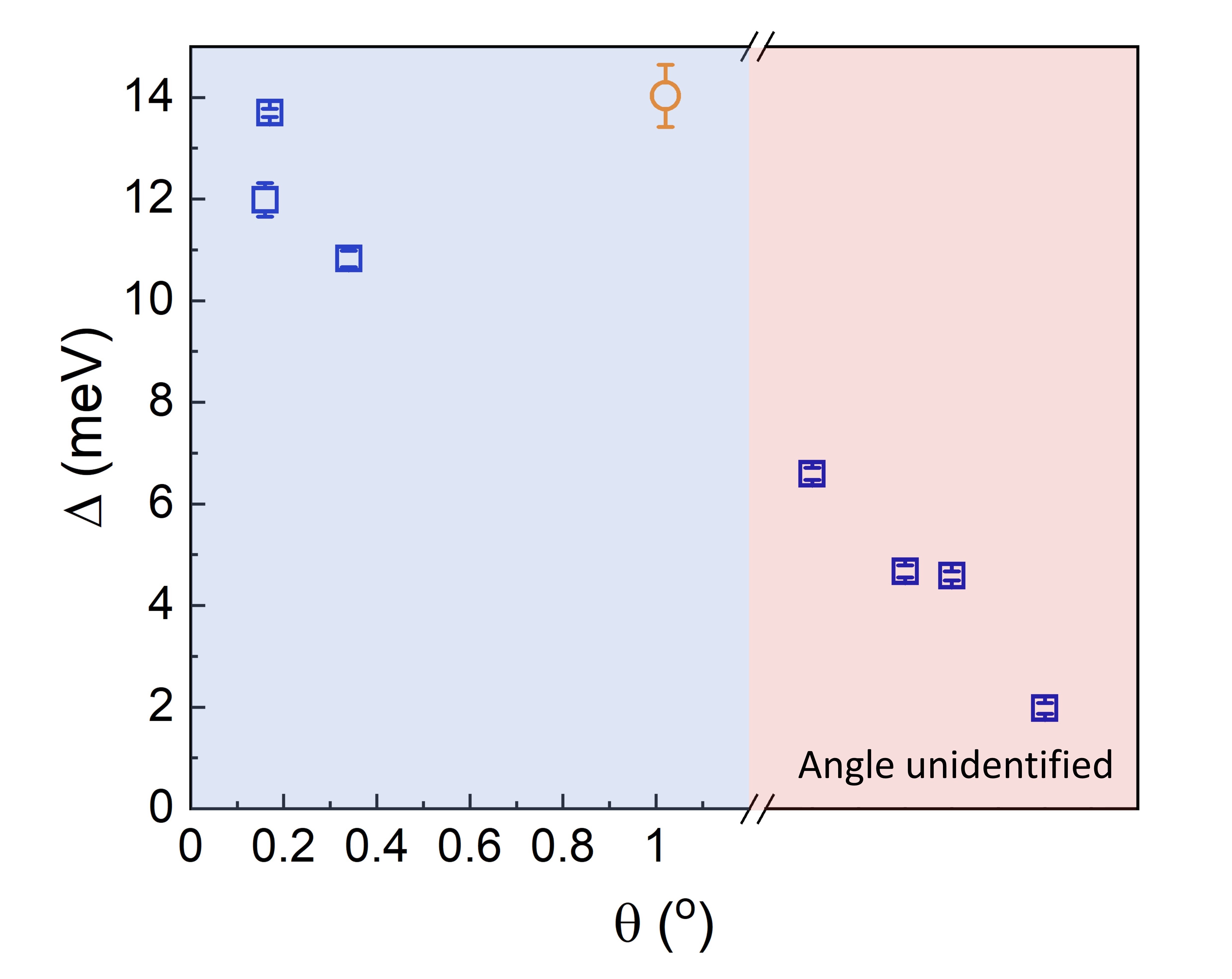}
\caption{\label{fig4}Summary of the extracted gap sizes $\Delta$ from devices with different moir\'e periods. The blue shaded area includes devices where the twist angle between graphene and hBN can be extracted from the satellite peak position in transport\cite{supplemental}. The pink shaded area includes devices with relatively large twist angles, which cannot be determined due to limited gate voltage range in experiment. Error bars show the standard deviations of multiple measurements at different magnetic fields. The orange circle data point is extracted via the first measurement scheme from device A, and other blue square data points are acquired via the second method.}
\end{figure}

In all of our measured devices with different twist angles, the gap sizes at the CNP ($\sim$14 meV at maximum) are all significantly smaller than the sizes reported from previous transport measurements ($\sim$30 -- 40 meV) \cite{hunt2013massive,woods2014commensurate,wang2015evidence,ribeiro2018twistable,finney2019tunable}. This observation is contrary to experience: the transport measured gap is usually smaller than the true band gap due to the existence of in-gap states. However, such kind of disagreement between thermal activation and spectroscopic measurement is not only seen in the case of graphene/hBN superlattice. For example, in the recent study of the moir\'e gap in twisted bilayer graphene, the transport measurement extracted a $\sim$32 meV gap at full filling of the valence band side through thermal activation measurement \cite{cao2018correlated}, while the scanning tunneling spectroscopy (STS) study extracted a $\sim$15 meV gap \cite{xie2019spectroscopic}. On the conduction band side, the transport measurement extracted a $\sim$40 meV gap\cite{cao2018correlated} while the STS study extracted a $\sim$25 meV gap\cite{choi2019electronic}. We believe the value extracted from our measurement is closer to the true gap size as we performed spectroscopy, where the energy scale can be directly read from our data. We noted that a recent planer tunneling spectroscopy measurement observed a gap size at CNP very close to ours\cite{kim2018accurate}.

We would also like to discuss the plausibility of possible excitonic effects in explaining the discrepancy between our experiment and previous transport experiments. In semiconductors, excitations of bound exciton states result in an optical band gap that is smaller than the true single-particle band gap. However, we believe that the excitonic effect in monolayer graphene/hBN gap is negligible. The strength of excitonic effects is usually proportional to the flatness of the band, which can be quantified by the effective mass of the carriers. The effective mass in gapped monolayer graphene is related to the gap size as $\Delta=2m^{*}v_F^2$. With $\Delta\sim 14$ meV and $v_F\sim$1.21$\times 10^6$ m/s extracted from device A, $m^*\sim 0.84\times 10^{-3}m_0$, where $m_0$ is the bare electron mass. This is much smaller than that in bilayer graphene ($m^*\sim0.04m_0$ \cite{zou2011effective}). So, excitonic effects and exciton binding energy in this gap is expected to be much weaker in monolayer graphene compared to those in bilayer graphene. Experimentally, we only observed a step-function like photocurrent spectrum at zero magnetic field as shown in Fig. 2(a), in contrast to sharp exciton peaks as shown in bilayer graphene in similar sample configurations\cite{ju2017tunable}. In bilayer graphene, the exciton binding energy at a band gap of 30 meV is only $\sim$5 meV\cite{ju2017tunable}. Since the excitonic effect should be weaker in monolayer graphene than in bilayer graphene, the exciton binding energy is likely to be smaller than 5 meV--too small to explain the difference between 30--40 meV in previous measurements\cite{hunt2013massive,woods2014commensurate,wang2015evidence,ribeiro2018twistable,finney2019tunable} and 14 meV in our experiment. At the same time, in our second approach to extract the band gap size, similar excitonic effects on inter-LL transitions should happen for both transitions at $\nu$ = +2 and -2. By subtracting their transition energies, the effect of exciton binding energy on determining the single-particle band gap is presumably cancelled out.

Theoretically, in the single-particle picture for a superlattice with 0$^{\circ}$ twist angle, a continuum model gives a $\sim$2 meV gap at the CNP\cite{moon2014electronic}. A previous DFT calculation obtained a $\sim$4 meV gap \cite{sachs2011adhesion}, which is on the same order of magnitude. Additional lattice relaxation enlarges the gap size to $\sim$7 meV \cite{jung2015origin,jung2017moire}, which is around half the value of our results. In addition to the single-particle model, many-body electron interaction effects may also enhance the gap size. Interacting Dirac particles in graphene are subjected to the A-B sublattice-dependent potential in the superlattice. The interactions can create sublattice correlation which can further amplify the potential, thus increase the gap\cite{song2013electron}. Theoretical calculations obtain $\sim$ 20 meV gap with both lattice relaxation and interaction effects taken into consideration\cite{jung2015origin}. This value is very close to our results.

In summary, we employed FTIR photocurrent spectroscopy to measure high-quality devices of hBN-encapsulated monolayer graphene. We observed a band gap of less than 14 meV at nearly zero-degree twist angle. This quantitative information should aid in the better understanding of the physics of graphene-based 2D moir\'e superlattices. Particularly, it allows for more accurate modeling of correlated electron physics and topological phenomena in twisted bilayer graphene\cite{cao2018correlated,cao2018unconventional,yankowitz2019tuning,lu2019superconductors,serlin2020intrinsic,sharpe2019emergent}, ABC trilayer graphene/hBN\cite{chen2019evidence,chen2019signatures,chen2020tunable}, twisted double bilayer graphene\cite{liu2020tunable,cao2020tunable} and twisted mono/bilayer graphene\cite{chen2020electrically,polshyn2020electrical}. The (magneto-) FTIR photocurrent spectroscopy could also be further extended to spectroscopic measurement of low energy electronic band structures of 2D moir\'e superlattices, as well as symmetry broken ground states in the quantum Hall regime\cite{young2012spin,young2014tunable,zibrov2018even,hunt2013massive,wang2015evidence,dean2013hofstadter,herbut2007theory,nomura2009field,kharitonov2012phase,kharitonov2012canted,alicea2006graphene}.\\[2mm]

We acknowledge discussions with R. Ashoori, L. Fu, and L. Levitov. This work was supported by the STC Center for Integrated Quantum Materials, NSF Grant No. DMR-1231319. T. H acknowledges the support of the Sloan Fund (No. 2695400) from the School of Science at MIT. T.H. and L.J. acknowledge the support from Materials Research Laboratory (MRL) at MIT. L. W. acknowledges the National Natural Science
Foundation of China (No. 12074173). K. W. and T. T. acknowledge support from the Elemental Strategy Initiative conducted by the MEXT, Japan ,Grant Number JPMXP0112101001,  JSPS
KAKENHI Grant Number JP20H00354 and the CREST(JPMJCR15F3), JST. The device fabrication in this work is supported by the Skolkovo Institute of Science and Technology as part of the MIT Skoltech Program, and was performed in part at the Center for Nanoscale Systems (CNS), a member of the National Nanotechnology Coordinated Infrastructure Network (NNCI), which is supported by the National Science Foundation under NSF award No. 1541959. CNS is part of Harvard University.
\bibliography{reference}
\end{document}